\documentclass[10pt]{article}
\usepackage{axodraw}
\begin{document}
\hoffset -1.7cm

\setlength{\hsize}{14.5cm}
\setlength{\vsize}{20cm}
\renewcommand{\baselinestretch}{1.2}

\title{\bf Minimal scalar sector of 3-3-1 models without exotic electric
charges}
\author{William A. Ponce and Yithsbey Giraldo\\
Instituto de F\'\i sica, Universidad de Antioquia,\\
A.A. 1226, Medell\'\i n, Colombia.
\and
Luis A. S\'anchez\\
Escuela de  F\'\i sica, Universidad Nacional de Colombia\\
A.A. 3840, Medell\'\i n, Colombia.}

\maketitle

\begin{abstract}
We study the minimal set of higgs scalars, for models based on the local
gauge group $SU(3)_c\otimes SU(3)_L\otimes U(1)_X$ which do not contain
particles with exotic electric charges. We show that only two higgs
$SU(3)_L$ triplets are needed in order to properly break the symmetry. The
exact tree-level scalar mass matrices resulting from symmetry breaking are
calculated at the minimum of the most general scalar potential, and the
gauge bosons are obtained, together with their coupling to the
physical scalar fields. We show how the scalar sector introduced is
enough to produce masses for fermions in a particular model which is an
$E_6$ subgroup. By using experimental results we constraint the
scale of new physics to be above 1.3 TeV.
\end{abstract}

\large

\section{Introduction}
The Standard Model (SM) based on the local gauge group  $SU(3)_c\otimes
SU(2)_L\otimes U(1)_Y$ \cite{sm} can be extended in several different ways:
first by adding new fermion fields (adding a right-handed neutrino field
constitute its simplest extension and has profound consequences, as for
example the implementation of the see-saw mechanism, and the enlarging of
the possible number of local gauge abelian symmetries that can be gauged
simultaneously); second, by augmenting the scalar sector to more than one
higgs representation, and third by enlarging the local gauge group. In
this last direction $SU(3)_L\otimes U(1)_X$ as a flavor group has been
studied previously by many authors in the literature \cite{pf}-\cite{pgs}
who have explored possible fermion and higgs-boson representation
assignments. From now on, models based on the local gauge group
$SU(3)_c\otimes SU(3)_L\otimes U(1)_X$ are going to be called 331 models.

There are in the literature several 331 models; the most popular one, the
Pleitez-Frampton model \cite{pf}, is far from being the simplest
construction. Not only its scalar sector is quite complicated and messy
(three triplets and one sextet \cite{scalar}), but its physical spectrum is
plagued with particles with exotic electric charges, namely: double charged
gauge and higgs bosons and exotic quarks with electric charges $5/3$ and
$-4/3$. Other 331 models in the literature are just introduced or merely
sketched in a few papers \cite{valle, ozer, spm, mps}, with a detailed
phenomenological analysis of them still lacking. In particular, there is
not published papers related to the study of the scalar sector for those
other models.

All possible 331 models without exotic electric charges in their gauge
boson sector and in their spin 1/2 fermion content are presented in
Ref. \cite{pfs}, where it is shown that there are just a few
anomaly free models for one or three families which share in common all
of them the same gauge-boson content and, as we are going to show next,
they may share a common scalar sector too. This scalar sector does
not contain particles with exotic electric charges either.

Our study is organized as follows: in section two we analyze the electric
charge operator in the context of 331 models. In section three we present
all the possible 331 models without exotic electric charges for one and
three families. In section four we study the (common) scalar sector for
all those models, including the analysis of its mass spectrum. In section
five we analyze the gauge boson structure common to all the models
considered. In section six we present the couplings between the neutral
scalar fields in the model and the SM gauge bosons, and in the last
section we present our conclusions. An appendix at the end shows how the
higgs scalars used to break the symmetry, can also be used to produce a
consistent mass spectrum for the fermion fields, in the particular model
which is an $E_6$ subgroup \cite{spm}.

\section{Charge content of 331 models}
In what follows we assume that the electroweak group is $SU(3)_L\otimes
U(1)_X\supset SU(2)_L\otimes U(1)_Y$. We also assume that the left-handed
quarks (color triplets) and left-handed leptons (color singlets) transform
under the two fundamental representations of $SU(3)_L$ (the 3 and $3^*$)
and that $SU(3)_c$ is vectorlike as in the SM.

The most general electric charge operator in $SU(3)_L\otimes U(1)_X$
is a linear combination of the three diagonal generators of the gauge
group
\begin{equation}\label{ch}
Q=aT_{3L}+\frac{2}{\sqrt{3}}bT_{8L}+XI_3,
\end{equation}
where $T_{iL}=\lambda_{iL}/2$, being $\lambda_{iL}$ the Gell-Mann matrices
for $SU(3)_L$ normalized as {\bf Tr.}$(\lambda_i\lambda_j)=2\delta_{ij}$,
$I_3=Dg(1,1,1)$ is the diagonal $3\times 3$ unit matrix,
and $a$ and $b$ are arbitrary parameters to be determined anon. The $X$
values are fixed by anomaly cancelation \cite{pfs, pgs} and an eventual
coefficient for $XI_3$ can be absorbed in the hypercharge definition.

If we assume that the usual isospin $SU(2)_L$ of the SM is
such that $SU(2)_L\subset SU(3)_L$, then $a=1$ and we have just a one
parameter set of models, all of them characterized by the value of $b$.
So, Eq.(\ref{ch}) allows for an infinite number of models in the context
of the 331 gauge structure, each one associated to a particular value of
the parameter $b$, with characteristic signatures that make each one quite
different from each other.

There are a total of 17 gauge bosons in the gauge group under
consideration, they are: one gauge field $B^\mu$ associated with $U(1)_X$,
the 8 gluon fields associated with $SU(3)_c$ which remain massless after
breaking the symmetry, and other 8 associated with $SU(3)_L$ and that we
may write in the following way:

\[{1\over 2}\lambda_{\alpha L} A_\mu^\alpha={1\over \sqrt{2}}\left(
\begin{array}{ccc}D^0_{1\mu} & W^+_\mu & K^{(1/2+b)}_\mu \\ W^-_\mu &
D^0_{2\mu} &  K^{-(1/2-b)}_\mu \\
K^{-(1/2+b)}_\mu & \bar{K}^{1/2-b}_\mu & D^0_{3\mu} \end{array}\right) \]
where $D^0_{1\mu}=A_{3\mu}/\sqrt{2}+A_{8\mu}/\sqrt{6},\;
D^0_{2\mu}=-A_{3\mu/}\sqrt{2}+A_{8\mu}/\sqrt{6}$,
and $D^0_{3\mu}=-2A_{8\mu/}\sqrt{6}$.
The upper indices of the gauge bosons in the former expression stand for
the electric charge of the corresponding particle, some of them functions
of the $b$ parameter as they should be \cite{pgs}. Notice that the gauge
bosons have integer electric charges only for $b=\pm 1/2,\;\pm 3/2,\;\pm
5/2,...,\pm (2n+1)/2,\; n=0, 1,2,3,...$. A deeper analysis shows that each
negative $b$ value can be related to the positive one just by taking the
complex conjugate in the covariant derivative, which in turn is equivalent
to replace $3\leftrightarrow 3^*$ in the fermion content of each
particular model.

Our first conclusion is thus that if we want to avoid exotic electric
charges in the gauge sector of our theory, then $b$ must be equal to 1/2,
which is also the condition for excluding exotic electric charges in the
fermion sector \cite{pgs}.

Now, contrary to the SM where only the abelian $U(1)_Y$ factor is
anomalous, in the 331 theory both, $SU(3)_L$ and $U(1)_X$ are anomalous
($SU(3)_c$ is vectorlike as in the SM). So, special combination of
multiplets must be used in each particular model in order to cancel the
several possible anomalies, and end with physical acceptable models. The
triangle anomalies we must take care of are: $[SU(3)_L]^3,\;
[SU(3)_c]^2U(1)_X, \; [SU(3)_L]^2U(1)_X,\; [grav]^2U(1)_X$ (the
gravitational anomaly), and $[U(1)_X]^3$.

In order to present specific examples let us see how the charge operator
in Eq.(\ref{ch}) acts on the representations 3 and $3^*$ of $SU(3)_L$:

\[Q[3]=Dg.({1 \over{2}}+{b \over{3}}+X, -{1 \over{2}}+{b \over{3}}+X,
-{2b \over{3}}+X)\]
\[Q[3^*]=Dg.(-{1 \over{2}}-{b \over{3}}+X, {1 \over{2}}-{b \over{3}}+X,
{2b \over{3}}+X).\]

Notice from this expressions that, if we accommodate the known left-handed
quark and lepton isodoublets in the two upper components of 3 and $3^*$
(or $3^*$ and 3), and forbid the presence of exotic electric charges in
the possible models, then the electric charge of the third component in
those representations must be equal either to the charge of the first or
second component, which in turn implies $b=\pm 1/2$. Since the negative
value is equivalent to the positive one, $b=1/2$ is a necessary and
sufficient condition in order to exclude exotic electric charges in the
fermion sector too.

As an example of the former discussion let us take $b=3/2$, then
$Q[3]=Dg.(1+X,X,X-1)$ and $Q[3^*]=Dg.(X-1,X,1+X)$. Then the following
multiplets are associated with the respective $(SU(3)_c,SU(3)_L,U(1)_X)$
quantum numbers: $(e^-,\nu_e,e^+)^T_L\sim (1,3^*,0);\; (u,d,j)^T_L\sim
(3,3,-1/3)$ and $(d,u,k)^T_L\sim (3,3^*,2/3)$, where $j$ and $k$ are
isosinglet exotic quarks of electric charges $-4/3$ and $5/3$
respectively. This multiplet structure is the basis of the
Pleitez-Frampton model \cite{pf} for which the anomaly-free arrangement for
the three families is given by:
\begin{eqnarray*}
\psi_L^a&=&(e^a,\nu^a,e^{ca})^T_L\sim (1,3^*,0),\\
q_L^i&=&(u^i,d^i,j^i)^T_L\sim (3,3,-1/3),\\
q_L^1&=&(d^1,u^1,k)^T_L\sim (3,3^*,2/3),\\
u^{ca}_L&\sim&(3^*,1,-2/3),\;\; d^{ca}_L\sim(3^*,1,1/3),\\
k^a_L&\sim&(3^*,1,-5/3),\;\; j^{ci}_L\sim (3^*,1,-4/3),
\end{eqnarray*}
where the upper $c$ symbol stands for charge conjugation, $a=1,2,3$ is a
family index and $i=2,3$ is related to two of the three families (in the
331 basis). As can be seen, there are six triplets of $SU(3)_L$ and six
anti-triplets, which ensures cancelation of the $[SU(3)_L]^3$ anomaly. A
power counting shows that the other four anomalies also vanish.

\section{Models without exotic electric charges}
As discussed before, after fixing $a=1$, the value $b=1/2$ is a necessary
condition in order to avoid particles with exotic electric charges in
models based on the $SU(3)_c\otimes SU(3)_L\otimes U(1)_X$ gauge
structure. For that particular value let us start first defining the
following closed set of fermions (closed in the sense that they include
the antiparticles of the charged particles):\\
\begin{itemize}
\item $S_1=[(\nu_\alpha,\alpha^-,E_\alpha^-);\alpha^+;E^+_\alpha]$ with
quantum numbers [$(1,3,-2/3)$; $(1,1,1)$; $(1,1,1)$].
\item $S_2=[(\alpha^-,\nu_\alpha,N^0_\alpha);\alpha^+]$ with quantum
numbers [$(1,3^*,-1/3)$; $(1,1,1)$].
\item $S_3=[(d,u,U);d^c;u^c;U^c]$ with quantum numbers
$(3,3^*,1/3)$; $(3^*,1,1/3)$; $(3^*,1,-2/3)$ and $(3^*,1,-2/3)$, respectively.
\item $S_4=[(u,d,D);u^c;d^c;D^c]$ with quantum numbers
$(3,3,0)$; $(3^*,1,-2/3)$; $(3^*,1,1/3)$ and $(3^*,1,1/3)$, respectively.
\item $S_5=[(e^-,\nu_e,N^0_1);(E^-,N^0_2,N^0_3);(N^0_4,E^+,e^+)]$ with
quantum numbers $(1,3^*,-1/3)$; $(1,3^*,-1/3)$ and $(1,3^*,2/3)$,
respectively.
\item $S_6=[(\nu_e,e^-,E^-_1);(E_2^+,N_1^0,N_2^0);(N_3^0,E_2^-,E_3^-);e^+;
E_1^+;E_3^+]$ with quantum numbers
[$(1,3,-2/3)$; $(1,3,1/3)$; $(1,3,-2/3)$; $(1,1,1)$; $(1,1,1)$; $(1,1,1)$].
\end{itemize}
Where the quantum numbers in parenthesis refer to
$(SU(3)_c,SU(3)_L,U(1)_X)$ representations.

The several anomalies for the former six sets are presented in the
following Table.

\begin{center}

TABLE I. Anomalies for $S_i$.

\begin{tabular}{||l||c|c|c|c|c|c||}\hline\hline
Anomalies           & $S_1$& $S_2$& $S_3$& $S_4$& $S_5$ & $S_6$ \\
\hline\hline
$[SU(3)_c]^2U(1)_X$ & 0    &  0    &  0    &  0   &  0  & 0 \\
$[SU(3)_L]^2U(1)_X$ & $-2/3$ & $ -1/3$ &  1    &  0   &  0  & $-1$ \\
$[grav]^2U(1)_X$     & 0    &  0    &  0    &  0   &  0  & 0 \\
$[U(1)_X]^3$        & 10/9 & 8/9   & $-12/9$ & $-6/9$ & 6/9 & 12/9 \\
$[SU(3)_L]^3$       & 1    & $-1$  & $-3$    & 3      & $-3$ & 3 \\
\hline\hline
\end{tabular}
\end{center}

Table I allows us to build anomaly-free models without exotic electric
charges, for one, two, three, four or more families. Let us extract out
of the Table the possible models for one and three families:
\subsection{One family models}
There are just two anomaly-free one family structures that can be
extracted from the Table. They are:\\
{\bf Model A}: $(S_4+S_5)$. This models is associated with an $E_6$
subgroup and has been partially analyzed in Ref. \cite{spm}.
(see also the appendix at the end of this paper). \\
{\bf Model B}: $(S_3+S_6)$. This models is associated with an
$SU(6)_L\otimes U(1)_X$ subgroup and has been partially analyzed in
Ref. \cite{mps}

\subsection{Three family models}
{\bf Model C}: $(3S_2+S_3+2S_4)$.
This model deals with the following multiplets associated
with the given quantum numbers: $(u,d,D)^T_L\sim (3,3,0),\;
(e^-,\nu_e,N^0)^T_L\sim (1,3^*,-1/3)$ and $(d,u,U)^T_L\sim
(3,3^*,1/3)$, where $D$ and $U$ are exotic quarks with electric charges
$-1/3$ and 2/3 respectively. With such a gauge structure the three
family anomaly-free model is given by:
\begin{eqnarray*}
\psi_L^{'a}&=&(e^{-a},\nu^a,N^{0a})^T_L\sim (1,3^*,-1/3),\\
e^{+a}_L &\sim& (1,1,1),\\
q_L^{'i}&=&(u^i,d^i,D^i)^T_L\sim (3,3,0),\\
q_L^{'1}&=&(d^1,u^1,U)^T_L\sim (3,3^*,1/3),\\
u^{ca}_L&\sim&(3^*,1,-2/3),\;\; d^{ca}_L\sim (3^*,1,1/3),\\
U^c_L&\sim&(3^*,1,-2/3),\;\; D^{ci}_L\sim (3^*,1,1/3),
\end{eqnarray*}
where $a=1,2,3$ is a family index and $i=1,2$ is related to two of the
three families. This models has been analyzed in the literature in
Ref. \cite{valle}. If needed, this model can be augmented with an
undetermined number of neutral Weyl states $N^{0j}_L\sim (1,1,0),\;
j=1,2,...$ without violating the anomaly cancelation.

{\bf Model D}: $(3S_1+2S_3+S_4)$. It makes use of the same multiplets used
in the previous model arranged in a different way, plus a new lepton
multiplet $(\nu_e,e^-,E^-)^T_L\sim (1,3,-2/3)$. The family
structure of this new anomaly-free model is given by:

\begin{eqnarray*}
\psi_L^{''a}&=&(\nu^a,e^a,E^{a})^T_L\sim (1,3,-2/3),\\
e^{ca}_L&\sim&(1,1,1),\;\; E^{ca}_L\sim (1,1,1),\\
q_L^{''1}&=&(u^1,d^1,D)^T_L\sim (3,3,0),\\
q_L^{''i}&=&(d^i,u^i,U^i)^T_L\sim (3,3^*,1/3),\\
u^{ca}_L&\sim&(3^*,1,-2/3),\;\; d^{ca}_L\sim (3^*,1,1/3),\\
D^c_L&\sim&(3^*,1,1/3),\;\; U^{ci}_L\sim (3^*,1,2/3).
\end{eqnarray*}
This model has been analyzed in the literature in Ref. \cite{ozer}.

{\bf Model E}: $(S_1+S_2+S_3+2S_4+S_5)$.
{\bf Model F}: $(S_1+S_2+2S_3+S_4+S_6)$.

Besides the former four three family models, other four, carbon copy of
the two one family models {\bf A,B} can also be constructed. They are:

{\bf Model G}: $3(S_4+S_5)$.
{\bf Model H}: $3(S_3+S_6)$.
{\bf Model I}: $2(S_4+S_5)+(S_3+S_6)$.
{\bf Model J}: $(S_4+S_5)+2(S_3+S_6)$.

For a total of eight different three-family models, each one with a
different fermion field content. Notice in particular that in models
{\bf E} and {\bf F} each one of the three families is treated differently.
As far as we know the last six models have not been studied in the
literature so far.

If we wish we may construct also two, four, five, etc. family models (a
two family model is given for example by $(S_1+S_2+S_3+S_4))$, but we
believe all those models are not realistic at all.

\section{The scalar sector}
If we pretend to use the simplest $SU(3)_L$ representations in order to
break the symmetry, at least two complex scalar triplets, equivalent to
twelve real scalar fields, are required. For $b=1/2$ there are just
two higgs scalars (together with their complex conjugates) which may
develop nonzero Vacuum Expectation Values (VEV); they are
$\phi_1(1,3^*,-1/3)^T=(\phi_1^-,\phi_1^{'0},\phi_1^0)$ with VEV
$\langle\phi_1\rangle^T=(0,v_1,V)$ and
$\phi_2(1,3^*,2/3)^T=(\phi_2^0,\phi_2^+,\phi_2^{'+})$ with VEV
$\langle\phi_2\rangle^T=(v_2,0,0)$. As we will see ahead, to reach
consistency with phenomenology we must have the hierarchy
$V>v_1\sim v_2$.

Our aim is to break the symmetry in one single step

\[SU(3)_c\otimes SU(3)_L\otimes U(1)_X\longrightarrow SU(3)_c\otimes
U(1)_Q\]

\noindent
which implies the existence of eight Goldstone bosons included in the
scalar sector of the theory \cite{goldstone}. For the sake of simplicity we
assume that the VEV are real. This means that the CP violation through the
scalar exchange is not considered in this work. Now, for convenience in
reading we rewrite the expansion of the scalar fields which acquire VEV
as:

\begin{equation} \label{neutras}
\phi_1^0=V+\frac{H^0_{\phi_1}+iA^0_{\phi_1}}{\sqrt{2}} \hspace{1cm}
\phi_1^{'0}=v_1+\frac{H^{'0}_{\phi_1}+iA^{'0}_{\phi_1}}{\sqrt{2}}
\hspace{1cm}
\phi_2^0=v_2+\frac{H^0_{\phi_2}+iA^0_{\phi_2}}{\sqrt{2}} .
\end{equation}

In the literature, a real part $H$ is called a CP-even scalar and an
imaginary part $A$ a CP-odd scalar or pseudoscalar field.

Now, the most general potential which includes $\phi_1$ and $\phi_2$ can
then be written in the following form:
\begin{equation}\label{vphi}
V(\phi_1,\phi_2)=
\mu^2_1\phi_1^\dagger\phi_1 + \mu^2_2\phi_2^\dagger\phi_2 +
\lambda_1(\phi_1^\dagger\phi_1)^2 +\lambda_2(\phi_2^\dagger\phi_2)^2 +
\lambda_3(\phi_1^\dagger\phi_1)(\phi_2^\dagger\phi_2) +
\lambda_4(\phi_1^\dagger\phi_2)(\phi_2^\dagger\phi_1).
\end{equation}

Requiring that in the shifted potential $V(\phi_1,\phi_2)$, the linear
terms in fields must be absent, we get in the tree-level approximation the
following constraint equations:
\begin{eqnarray}\nonumber
\mu^2_1+2\lambda_1(v_1^2+V^2)+\lambda_3v_2^2&=&0\\ \label{minimos}
\mu^2_2+\lambda_3(v_1^2+V^2)+2\lambda_2v_2^2&=&0.
\end{eqnarray}
The analysis to the former equations shows that they are related
to a minimum of the scalar potential with the value
\begin{equation}\label{min}
V_{min}=-v_2^4\lambda_2-(v_1^2+V^2)[(v_1^2+V^2)\lambda_1+v_2^2\lambda_3]
=V(v_1,v_2,V),
\end{equation}
where $V(v_1=0,v_2,V)>V(v_1\neq 0,v_2,V)$, implying that $v_1\neq 0$ is
preferred.

Substituting Eqs.(\ref{neutras}) and (\ref{minimos}) in Eq.(\ref{vphi}) we
get the following mass matrices:

\subsection{Spectrum in the scalar neutral sector}
In the $(H^0_{\phi_1},H^0_{\phi_2},H^{'0}_{\phi_1})$ basis, the square
mass matrix can be calculated using
$M_{ij}^2=2\frac{\partial^2V(\phi_1\phi_2)}{\partial H^0_{\phi_i}\partial
H^0_{\phi_j}}$ . After imposing the constraints in Eq.(\ref{minimos}) we
get:
\begin{equation} \label{matrix}
M_H^2=2\left(\begin{array}{ccc}
2\lambda_1V^2 & \lambda_3v_2V   & 2\lambda_1v_1V \\
\lambda_3v_2V & 2\lambda_2v_2^2 & \lambda_3v_1v_2 \\
2\lambda_1v_1V & \lambda_3v_1v_2 & 2\lambda_1v_1^2
\end{array}\right),
\end{equation}
which has zero determinant, providing us with a Goldstone boson $G_1$ and
two physical massive neutral scalar fields $H_1$ and $H_2$ with masses
\begin{equation} \nonumber
M^2_{H_1,H_2}=2(v_1^2+V^2)\lambda_1+2v_2^2\lambda_2\pm
2\sqrt{[(v_1^2+V^2)\lambda_1 +
v_2^2\lambda_2]^2+v_2^2(v_1^2+V^2)(\lambda_3^2-4\lambda_1\lambda_2)},
\end{equation}
where real lambdas produce positive masses for the scalars only
if $\lambda_1>0$ and  $4\lambda_1\lambda_2 >\lambda_3^2$ (which implies
$\lambda_2>0$).

We may see from the former equations that in the limit $V>v_1\sim v_2$,
and for lambdas of order one, there is a neutral higgs scalar with a mass
of order $V$ and other one with a mass of the order of $v_1\sim v_2$,
which may be identified with the SM scalar as we will see ahead.

The physical fields are related to the scalars in the weak basis by
the lineal transformation:
\begin{equation}
\left( \begin{array}{ccc}
H_{\phi^1}^0 \\ H_{\phi^2}^0 \\ H_{\phi^1}^{'0} \end{array}\right) =
\left( \begin{array}{ccc}
\frac{v_2V}{S_1} & \frac{v_2V}{S_2} &  -\frac{v_1}{\sqrt{v_1^2+V^2}} \\
\frac{M_{H_1}^2-4(v_1^2+V^2)\lambda_1}{2S_1\lambda_3} &
-\frac{(M_{H_1}^2-4v_2^2\lambda_2)}{2S_2\lambda_3} & 0 \\
\frac{v_1v_2}{S_1} & \frac{v_1v_2}{S_2} & \frac{V}{\sqrt{v_1^2+V^2}}
\end{array}\right) \left( \begin{array}{ccc}
H_1 \\ H_2 \\ G_1 \end{array}\right),
\nonumber\end{equation}
where we have defined
$S_1=\sqrt{v_2^2(v_1^2+V^2)+(M_{H_1}^2-4(v_1^2+V^2)\lambda_1)^2/4\lambda_3^2}$
and
$S_2=\sqrt{v_2^2(v_1^2+V^2)+(M_{H_1}^2-4v^2_2\lambda_2)^2/4\lambda_3^2}$.

\subsection{Spectrum in the pseudoscalar neutral sector}
The analysis shows that $V(\phi_1,\phi_2)$ in Eq.(\ref{vphi}), when
expanded around the most general vacuum given by Eqs.(\ref{neutras}) and
using the constraints in Eq.(\ref{minimos}), does not contain pseudoscalar
fields $A^0_{\phi_i}$. This allows us to identify another three Goldstone
bosons $G_2=A^0_{\phi_1},\; G_3=A^0_{\phi_2}$ and $G_4=A^{'0}_{\phi_1}$.

\subsection{Spectrum in the charged scalar sector}
In the basis $(\phi_1^+, \phi_2^+,\phi_2^{'+})$ the square mass matrix is
given by
\begin{equation}
M_+^2=2\lambda_4\left(\begin{array}{ccc}
v_2^2 & v_1v_2 & v_2V \\
v_1v_2 & v_1^2 & v_1V \\
v_2V & v_1V  & V^2 \end{array}\right),
\end{equation}
which has two eigenvalues equal to zero equivalent to four Goldstone
bosons $(G_5^\pm ,G_6^\pm)$ and two physical charged higgs scalars with
large masses given by $\lambda_4(v_1^2+v_2^2+V^2)$, with the new
constraint $\lambda_4>0$.

Our analysis shows that, after symmetry breaking, the original twelve
degrees of freedom in the scalar sector have become eight Goldstone bosons
(four electrically neutral and four charged), and four physical scalar
fields, two neutrals (one of them the SM higgs scalar) and two charged
ones. The eight Goldstone bosons must be swallow up by eight gauge fields
as we will see in the next section.

\section{The Gauge boson sector}
For $b=1/2$, the nine gauge bosons in $SU(3)_L\otimes U(1)_X$ when acting
on left-handed triplets can be arranged in the following convenient way:

\[{\cal A_\mu}={1 \over{2}}g\lambda_{\alpha L}A^\alpha_\mu + g'XB_\mu I_3=
{g \over{\sqrt{2}}}\left(\begin{array}{ccc}
Y^0_{1\mu} & W^+_\mu & K^+_\mu \\
W^-_\mu & Y^0_{2\mu} & K^0_\mu \\
K^-_\mu & \overline{K}^0_{\mu} & Y^0_{3\mu} \end{array}\right),\]

\noindent
where
$Y^0_{1\mu}=A_{3\mu}/\sqrt{2}+A_{8\mu}/\sqrt{6}+\sqrt{2}(g'/g)XB_\mu,
\;\;
Y^0_{2\mu}=-A_{3\mu}/\sqrt{2}+A_{8\mu}/\sqrt{6}+\sqrt{2}(g'/g)XB_\mu$,
and $Y^0_{3\mu}=-2A_{8\mu}/\sqrt{6}+\sqrt{2}(g'/g)XB_\mu; \;\; X$
being the hypercharge value of the given left-handed triplet
(for example $-1/3$ and 2/3 for $\phi_1$ and $\phi_2$ respectively).

After breaking the symmetry with $\langle\phi_i\rangle,\; i=1,2$, and
using for the covariant derivative for triplets $D^\mu=\partial^\mu
-i{\cal A^\mu}$, we get the following mass terms in the gauge boson
sector:

\subsection{Spectrum in the charged gauge boson sector}
In the basis $(K^\pm_\mu, W^\pm_\mu)$ the square mass matrix produced is
\begin{equation} M^2_{\pm}={g^2 \over{2}}\left(\begin{array}{cc}
(V^2+v_2^2) & v_1V \\ v_1V & (v_1^2 + v_2^2) \end{array}\right).
\end{equation}
The former symmetric matrix give us the masses
$M_{W'}^2=g^2v_2^2/2$ and $M_{K'}^2=g^2(v_1^2+v_2^2+V^2)/2$, related to
the physical fields $W^{'}_\mu=\eta (v_1K_\mu-VW_\mu),$ and $K^{'}_\mu =
\eta (VK_\mu+v_1W_\mu)$ associated with the known charged weak current
$W^{'\pm}_\mu$, and with a new one $K^{'\pm}_\mu$ predicted in the context
of this model ($\eta^{-2}=v_1^2+V^2$ is a normalization factor). From the
experimental value $M_{W'}=80.419 \pm 0.056$ GeV \cite{pdb} we obtain
$v_2\simeq 174$ GeV as in the SM.

\subsection{Spectrum in the neutral gauge boson sector}
For the five electrically neutral gauge bosons we get first, that the
imaginary part of $K^0_\mu=(K^0_{\mu R}+iK^0_{\mu I})/\sqrt{2}$ decouples
from the other four electrically
neutral gauge bosons, acquiring a mass $M^2_{K^0_I}=g^2(v_1^2+V^2)/2$.
Then,
in the basis $(B^\mu, A^\mu_3, A^\mu_8,K^{0\mu}_R)$, the following squared
mass matrix is obtained:

\begin{equation}\nonumber
M^2_{0}=\left(\begin{array}{cccc}
\frac{g^{'2}}{9}(v_1^2+V^2+4v^2_2) &
-\frac{gg'}{6}(v^2_1 + 2v^2_2) &
-{gg' \over{3\sqrt{3}}}(V^2+v_2^2-v^2_1/2)  & gg'v_1V/3  \\
-\frac{gg'}{6}(v^2_1 + 2v^2_2) & g^2(v^2_1+v^2_2)/4 &
{g^2 \over{4\sqrt{3}}}(v^2_2-v^2_1) & -g^2v_1V/4 \\
-{gg' \over{3\sqrt{3}}}(V^2 + v_2^2 -v^2_1/2)  &
{g^2 \over{4\sqrt{3}}}(v^2_2-v^2_1) &
{g^2 \over{12}}(v^2_1+v^2_2+4V^2) & -g^2v_1V/(4\sqrt{3}) \\
gg'v_1V/3 & -g^2v_1V/4 & -g^2v_1V/(4\sqrt{3}) &
g^2(v^2_1+V^2)/4 \end{array}\right)
\end{equation}

This matrix has determinant equal to zero which implies that
there is a zero eigenvalue associated to the photon field with eigenvector

\begin{equation}\label{foton}
A^\mu=S_W A_3^\mu + C_W\left[\frac{T_W}{\sqrt{3}}A_8^\mu+
(1-T_W^2/3)^{1/2}B^\mu\right],
\end{equation}
where $S_W=\sqrt{3}g'/\sqrt{3g^2+4g^{'2}}$ and $C_W$ are the sine and
cosine of the electroweak mixing angle ($T_W=S_W/C_W$). Orthogonal to the
photon field $A^\mu$ we may define other two fields
\begin{eqnarray}\nonumber
Z^\mu&=& C_W A_3^\mu - S_W\left[\frac{T_W}{\sqrt{3}}A_8^\mu+
(1-T_W^2/3)^{1/2}B^\mu\right]\\ \label{neutral}
Z'^\mu&=&-(1-T_W^2/3)^{1/2}A_8^\mu+\frac{T_W}{\sqrt{3}}B^\mu,
\end{eqnarray}
where $Z^\mu$ corresponds to the neutral current of the SM and $Z^{'\mu}$
is a new weak neutral current predicted for these models.

We may also identify the gauge boson $Y^\mu$
associated with the SM hypercharge in $U(1)_Y$ as:
\[Y^\mu=\left[\frac{T_W}{\sqrt{3}}A_8^\mu+ (1-T_W^2/3)^{1/2}B^\mu\right].\]

In the basis $(Z^{'\mu}, Z^\mu, K^{0\mu}_R)$ the mass matrix
for the neutral sector reduces to:
\begin{equation} \label{gauges}
\frac{g^2}{4C_W^2}\left( \begin{array}{ccc}
\delta^2(v_1^2C^2_{2W}+ v_2^2+4V^2C_W^4)  &
\delta (v_1^2C_{2W}-v_2^2) & \delta C_W v_1 V \\
\delta (v_1^2C_{2W}-v_2^2) & v_1^2+v_2^2 &-C_W v_1V  \\
\delta C_W v_1 V & -C_W v_1V  & C_W^2(v_1^2+V^2) \end{array}\right),
\end{equation}
where $C_{2W}=C^2_W-S^2_W$ and $\delta = (4C_W^2-1)^{-1/2}$. The
eigenvectors and eigenvalues of this matrix are the physical fields
and their masses. In the approximation $v_1=v_2\equiv v << V$ and using
$q\equiv v^2/V^2$ as an expansion parameter we get up, to first order in
$q$, the following eigenvalues:
\begin{eqnarray*}
M^2_{Z_1}&\approx& {1 \over{2}}g^2C_W^{-2}v^2(1-q T_W^4), \\
M^2_{Z_2}&\approx&
\frac{g^2V^2}{1+2C_{2W}}[1+C_{2W}-q(S^2_{2W}+C_{2W}^{-1})/2C_W^2],\\
M^2_{K'^0_R}&\approx& g^2V^2[1+q(1+C^{-1}_{2W})].
\end{eqnarray*}
So we have a neutral current associated to a gauge boson $Z_1^0$, related
to a mass scale $v\simeq 174$ GeV, which may be identified with the known
experimental neutral current as we will see in what follows, and two new
electrically neutral currents associated to a large mass scale $V>>v$.

The former is the way how the eight would be Goldstone bosons are absorbed
by the longitudinal components of the eight massive gauge bosons ($W'^\pm$,
$K'^\pm$, $K^0_I$, $K'^0_R$, $Z^0_1$ and $Z^0_2$) as expected.

>>>From the expressions for $M_{W'}$ and $M_{Z_1}$ we obtain $\rho_0 =
M_{W'}^2/(M_{Z_1}^2C_W^2)\approx 1+ T_W^4 q^2$, and the global fit for
$\rho_0=1.0012^{+0.0023}_{-0.0014}$ \cite{langa} provides us with the lower
limit $V\geq 1.3$ TeV (where we are using for $S^2_W=0.23113$ \cite{pdb}).
This result justifies the existence of the expansion parameter $q\leq
0.01$ which sets the scale of new physics, together with the hierarchy
$V>v_1\sim v_2$.

\section{Higgs-SM gauge boson couplings}

In order to identify the considered above Higgs bosons with the one in the
SM, in this section we present the couplings of the two neutral scalar
fields $H_1$ and $H_2$ from section 4 with the physical gauge bosons
$W^{'\pm}$ and $Z_1^0$; then we take the limit $V>>v = v_1 = v_2$ which
produces the couplings of the physical scalars $H_1$ and $H_2$ with
the SM gauge bosons $W^\pm$ and $Z^0$ .

When the algebra gets done we obtain the following trilinear couplings,
provided $\lambda_3<0$:
\begin{eqnarray*}
g(W'W'H_1)&=&\frac{g^2v_2[M_{H_1}^2
-4(v_1^2+V^2)\lambda_1]}{2\sqrt{2}S_1\lambda_3}
\stackrel{V>>v}{\longrightarrow}\frac{g^2v_2^2\lambda_3}{2\sqrt{2}\lambda_1V}\\
g(W'W'H_2)&=&\frac{g^2v_2(4v_2^2\lambda_2-M_{H_1}^2)}{2\sqrt{2}S_2\lambda_3}
\stackrel{V>>v}{\longrightarrow}\frac{g^2v_2}{\sqrt{2}}\\
g(Z_1^0Z_1^0H_1)&=&
\frac{g^2v_1}{S_1}\left[\frac{M_{H_1}^2-4(v_1^2+V^2)\lambda_1}
{4\sqrt{2}C_W^2\lambda_3}+q \frac{v_1^2(\lambda_1-\lambda_3S^2_W)}
{8\sqrt{2}C_W^2\lambda_1}+ \dots \right] \\
& &\stackrel{V>>v}{\longrightarrow}\frac{g^2v_1^2\lambda_3}{4\sqrt{2}
\lambda_1C_W^2V}\\
g(Z_1^0Z_1^0H_2) &=& \frac{g^2v_1V^2}{S_2}
\left[-\frac{\lambda_1}{\sqrt{2}\lambda_3 C_W^2}+q
\frac{4\lambda_1\lambda_2-\lambda_3^2+2\lambda_1^2(T_W^2C_W^{-2}-2)}
{4\sqrt{2}C_W^2\lambda_1\lambda_3}+ \dots \right] \\
& &\stackrel{V>>v}{\longrightarrow}\frac{g^2v_1}{2\sqrt{2}C_W^2},
\end{eqnarray*}

\noindent
where $g(W'W'H_i^0), \; i=1,2$ are exact expressions and
$g(Z_1^0Z_1^0H_i)$ are expansions in the parameter $q$ up to first order.

The quartic couplings are determined to be:
\begin{eqnarray*}
g(W'W'H_1H_1)&=&\frac{g^2[M_{H_1}^2
-4(v_1^2+V^2)\lambda_1]^2}{16S^2_1\lambda^2_3}
\stackrel{V>>v}{\longrightarrow}\frac{g^2v^2_2\lambda_3^2}{16\lambda_1^2V^2}\\
g(W'W'H_2H_2)&=&\frac{g^2(M^2_{H_1}-4v_2^2\lambda_2)^2}{16S_2^2\lambda_3^2}
\stackrel{V>>v}{\longrightarrow}\frac{g^2}{4}\\
g(Z_1^0Z_1^0H_1H_1)&=&
\frac{g^2}{S^2_1}\left[\frac{[M_{H_1}^2-4(v_1^2+V^2)\lambda_1]^2 }
{32C_W^2\lambda_3^2} +q \frac{2v_1^4\lambda_3^2
-S_W^2[M_{H_1}^2-4(v_1^2+V^2)\lambda_1]^2}
{64C_W^6\lambda_3^2}+ \dots \right] \\
& &\stackrel{V>>v}{\longrightarrow}\frac{g^2v^2_1\lambda_3^2}
{32\lambda_1^2 C_W^2V^2}\\
g(Z_1^0Z_1^0H_2H_2)&=&  \frac{g^2V^4}{S^2_2}\left[
\frac{\lambda_1^2}{2\lambda_3^2C_W^2}+q
\frac{\lambda_3^2-4\lambda_1\lambda_2+\lambda_1^2(4-C_W^{-2}T_W^2)}
{4C_W^2\lambda_3^2}+ \dots \right]\\
& & \stackrel{V>>v}{\longrightarrow}\frac{g^2}{8C_W^2}\\
\end{eqnarray*}
where as before $g(W'W'H_i^0H_i^0), \; i=1,2$ are exact expressions and
$g(Z_1^0Z_1^0H_i^0H_i^0)$ are
expansions in the parameter $q$ up to first order.

As can be seen, in the limit $V>v_1\sim v_2$ the couplings $g(W'W'H_2)$,
$g(Z_1^0Z_1^0H_2)$, $g(W'W'H_2H_2)$ and $g(Z_1^0Z_1^0H_2H_2)$ coincide
with those in the SM as far as $\lambda_3<0$. This gives additional
support to the hierarchy $V>v_1\sim v_2$.

Summarizing, from the couplings of the SM gauge bosons with the physical
Higgs scalars we can conclude, as anticipated before, that the scalar
$H_2$ can be identified with the SM neutral Higgs particle, and that
$Z_1^0$ can be associated with the known neutral current of the SM (more
support to this last statement is presented in the Appendix).

\section{Conclusions}
In this paper we have studied in detail the minimal scalar sector of some
models based on the local gauge group $SU(3)c\otimes SU(3)_L\otimes
U(1)_X$. By restricting the field representations to particles without
exotic electric charges we end up with ten different models, two one
family models and eight models for three families. The two one family
models are studied in the papers in Refs. \cite{spm, mps}, but enough
attention was not paid to the scalar sector in the analysis done.  As far
as we know, most of the three family models are new in the literature, but
models {\bf C} and {\bf D}, which has been partially analyzed in
Refs. \cite{valle} and \cite{ozer} respectively.

We have also considered the mass spectrum eigenstates of the most general
scalar potential specialized for the 331 models without exotic electric
charges, with two Higgs triplets with the most general VEV possible. It is
shown that in the considered models there is just one light neutral Higgs
scalar which can be identified with the SM Higgs scalar; there are besides
three more heavy scalars, one charged and its charge conjugate and one
extra neutral one.

The two triplets of $SU(3)_L$ scalars with the most general VEV possible
produces a consistent fermion mass spectrum at least for one of the models
in the literature and the scale of the new physics predicted by the class
of models analyzed in this paper lies above 1.3 TeV as shown in the main
text. This scale is consistent with the analysis done in other
papers \cite{spm, mps} using a different phenomenological analysis.

Finally notice that our analysis allows us to constraint all the
parameters in the scalar potential; that is, our model is a consistent one
as far as $\lambda_1>0, \; \lambda_2>0, \;
4\lambda_1\lambda_2>\lambda_3^2, \; \lambda_3<0$ and $\lambda_4>0$.

\section{Acknowledgments}
Work partially supported by Colciencias in Colombia and
by CODI in the U. de Antioquia. L.A. S\'anchez acknowledges
partial financial support from U. de Antioquia.

\appendix

\section{Appendix}
In this appendix we show how the fermion fields of a particular
model acquire masses with the Higgs scalars and VEV introduced in the main
text. The analysis is model dependent, so let us use the
one family model {\bf A}, for which the fermion multiplets are \cite{spm}
$\chi_L^T=(u,d,D)_L\sim (3,3,0)$; $u_L^c\sim (3^*,1,-2/3)$; $d_L^c\sim
(3^*,1,1/3)$, $D_L^c\sim (3^*,1,1/3)$;
$\psi_{1L}^T=(e^-,\nu_e, N_1^0)_L\sim(1,3^*,-1/3)$,
$\psi_{2L}^T=(E^-,N_2^0, N_3^0)_L\sim(1,3^*,-1/3)$, and
$\psi_{3L}^T=(N_4^0,E^+,e^+)_L\sim(1,3^*,2/3)$. As shown in
Ref. \cite{spm}, this structure corresponds to an $E_6$ subgroup.

\noindent
\subsection{Bare Masses for fermion fields}
The most general Yukawa Lagrangian that the Higgs scalars
in Section 4 produce for the fermion fields in this model, can be written
as ${\cal L}_Y={\cal L}^Q_Y+{\cal L}^l_Y$, with
\begin{eqnarray*}
{\cal L}_Y^Q &=& \chi^T_LC(h_u\phi_2u_L^c+h_D\phi_1D_L^c+
              h_{d}\phi_1d_L^c) + h.c., \\
{\cal L}^l_Y &=& \epsilon_{abc}[\psi_{1L}^aC(h_1\psi_{2L}^b\phi_2^c
             +h_2\psi_{3L}^b\phi_1^c) +
            \psi_{2L}^aCh_3\psi_{3L}^b\phi_1^c] + h.c. ,
\end{eqnarray*}
where $h_\eta , \; \eta=u,d,D,1,2,3$ are Yukawa couplings of
order one; $a,b,c$ are $SU(3)_L$ tensor indices and $C$ is the charge
conjugation operator.

Using for $\langle\phi_i\rangle, \; i=1,2$ the VEV in section 4 we get
$m_u=h_uv_2$ for the mass of the up type-quark and
for the down sector in the basis $(d,D)$ we get the mass matrix
\begin{equation}
M_d=\left(\begin{array}{cc}
          h_dv_1  &  h_dV  \\
          h_Dv_1 & h_DV \end{array}\right); \end{equation}
now, looking for the eigenvalues of $M_dM_d^\dagger$, we get
$\sqrt{(h_d^2+h_D^2)(v_1^2+V^2)}$ and zero.  Notice that
for $h_u=1$ and assuming for example that we are referring to the third
family, we obtain the correct mass for the top quark (remember from
Section 5 that $v_2\simeq 174$ GeV), the bottom quark
remains massless at zero level, and there is an exotic Bottom quark with a
very large mass. Since there is no way to distinguish between $d_L^c$ and
$D_L^c$ in the Yukawa Lagrangian it is just natural to impose the discrete
symmetry $h_d=h_D\equiv h$.

For the charged lepton sector the mass eigenvalues are 0 and
$\sqrt{(h_2^2+h_3^2)(v_1^2+V^2)}$, with similar consequences
as in the down quark sector, where again it is natural to impose the
symmetry $h_2=h_3\equiv h^\prime$.

The analysis of the neutral lepton sector is more elaborated; at zero
level and in the basis $(\nu,N_1,N_2,N_3,N_4)$ we get the mass
matrix:

\begin{equation}\nonumber
M_N=\left(\begin{array}{ccccc}
0 & 0 & 0 & h_1v_2 & -h_2V \\
0 & 0 & -h_1v_2 & 0 & h_2v_1 \\
0 & -h_1v_2 & 0 & 0 & -h_3V \\
h_1v_2 & 0 & 0 & 0 & h_3v_1 \\
-h_2V & h_2v_1 & -h_3V & h_3v_1 & 0 \end{array}\right),
\end{equation}
with eigenvalues $0, \;\pm h_1v_2$ and
$\pm\sqrt{h_1^2v_2^2+(h_2^2+h_3^2)(V^2+v_1^2)}$, which implies
a Majorana neutrino of zero mass and two Dirac neutral
particles with masses one of them at the electroweak mass scale and the
other one at the TeV scale.

So, at zero level the charged exotic particles get  large masses of
order $V>1.3$ TeV, the top quark and a Dirac neutral particle get masses
of order $v_2\sim 174$ GeV, there is a Dirac neutral particle with a
mass of order V, and the bottom quark, charged lepton and a Majorana
neutrino remain massless. In what follows we will see that they pick up a
radiative mass in the context of the model studied here.

\subsection{Currents}
The interactions among the charged gauge fields in Section 5 with the
fermions of {\bf Model A} are \cite{spm}:
\begin{eqnarray}\nonumber
H^{CC}&=&
\frac{g}{\sqrt{2}}[W_\mu^+(\bar{u}_L\gamma^\mu
d_L-\bar{\nu}_{eL}\gamma^\mu
e_L-\bar{N}^0_{2L}\gamma^\mu E_L^--\bar{E}^+_L\gamma^\mu N^0_{4L}) \\
\nonumber
&+&K_\mu^+(\bar{u}_L\gamma^\mu D_L-\bar{N}^0_{1L}\gamma^\mu
e_L-\bar{N}^0_{3L}\gamma^\mu E_L^--\bar{e}^+_L\gamma^\mu N^0_{4L}) \\
\nonumber
&+&K_\mu^0(\bar{d}_L\gamma^\mu D_L-\bar{N}^0_{1L}\gamma^\mu
\nu_{eL}-\bar{N}^0_{3L}\gamma^\mu N^0_{2L}-\bar{e}^+_L\gamma^\mu
E^+_L)] + h.c.,
\end{eqnarray}
where the first two terms constitute the charged weak current of the SM,
and $K^\pm, \; K^0$ and $\bar{K}^0$ are related to new charged
currents which violate weak isospin.

The algebra also shows that the neutral currents $J_\mu(EM),\; J_\mu(Z)$
and $J_\mu(Z')$, associated with the Hamiltonian
$H^0=eA^\mu J_\mu(EM)+\frac{g}{C_W}Z^\mu J_\mu(Z)
+\frac{g^\prime}{\sqrt{3}}  Z^{\prime\mu}J_\mu(Z^\prime)$ (where
$A_\mu$ is the photon field in Eq.(\ref{foton}) and
$Z_\mu$ and $Z^\prime_\mu$ are the neutral gauge bosons introduced in
Eq.(\ref{neutral})) are:
\begin{equation}
J_\mu(EM)=\frac{2}{3}\bar{u}\gamma_\mu u-\frac{1}{3}(\bar{d}\gamma_\mu d
+ \bar{D}\gamma_\mu D)-\bar{e}^-\gamma_\mu
e^--\bar{E}^-\gamma_\mu E^-,
\end{equation}
\begin{eqnarray}
J_\mu(Z)&=&J_{\mu ,L}(Z)-S_W^2J_\mu (EM),\\
J_\mu(Z^\prime)&=&T_WJ_\mu(EM)-J_{\mu ,L}(Z^\prime),
\end{eqnarray}
where $e=gS_W=g^\prime C_W\sqrt{(1-T_W^2/3)}>0$ is the electric charge,
$J_\mu(EM)$ is the (vectorlike) electromagnetic current,
and the two neutral left-handed currents are given by:
\begin{eqnarray} \nonumber
J_{\mu ,L}(Z)&=&\bar{u}_L\gamma_\mu u_L-\bar{d}_L\gamma_\mu d_L
+\bar{\nu}_{eL}\gamma_\mu \nu_{eL} -\bar{e}^-_L\gamma_\mu
e^-_L +\bar{N}^0_2\gamma_\mu N^0_2 - \bar{E}^-\gamma_\mu E^- \\
&=&\sum_f T_{3f}\bar{f}_L\gamma_\mu f_L,
\end{eqnarray}
\begin{eqnarray} \nonumber
J_{\mu ,L}(Z^\prime)&=&S_{2W}^{-1}(\bar{u}_L\gamma_\mu u_L
-\bar{e}^-_L\gamma_\mu e^-_L -\bar{E}^-_L\gamma_\mu E^-_L
-\bar{N}^0_{4L}\gamma_\mu N^0_{4L})\\ \nonumber
& & +T_{2W}^{-1}(\bar{d}_L\gamma_\mu d_L
-\bar{E}^+_L\gamma_\mu E^+_L -\bar{\nu}_{eL}\gamma_\mu \nu_{eL}
-\bar{N}^0_{2L}\gamma_\mu N^0_{2L})\\
& & -T_W^{-1}(\bar{D}_L\gamma_\mu D_L
-\bar{e}^+_L\gamma_\mu e^+_L -\bar{N}^0_{1L}\gamma_\mu N^0_{1L}
-\bar{N}^0_{3L}\gamma_\mu N^0_{3L}),
\end{eqnarray}
where $S_{2W}=2S_WC_W, \; T_{2W}=S_{2W}/C_{2W}, \;
\bar{N}^0_2\gamma_\mu N^0_2=
\bar{N}^0_{2L}\gamma_\mu N^0_{2L}+\bar{N}^0_{2R}\gamma_\mu N^0_{2R}=
\bar{N}^0_{2L}\gamma_\mu N^0_{2L}-\bar{N}^{0c}_{2L}\gamma_\mu N^{0c}_{2L}=
\bar{N}^0_{2L}\gamma_\mu N^0_{2L}-\bar{N}^0_{4L}\gamma_\mu N^0_{4L}$,
similarly $\bar{E}\gamma_\mu E=\bar{E}^-_L\gamma_\mu E^-_L -
\bar{E}^+_L\gamma_\mu E^+_L$ and $T_{3f}=Dg(1/2,-1/2,0)$ is the third
component of the weak isospin acting on the representation 3 of $SU(3)_L$
(the negative when acting on $\bar{3}$). Notice that $J_\mu (EM)$ and
$J_\mu (Z)$ are just the generalization of the electromagnetic and neutral
weak currents of the SM, as they should be, implying that $Z_\mu$ can
be identified as the neutral gauge boson of the SM.

\subsection{Radiative masses for fermion fields}
Using the currents in the previous section and the off diagonal entries in
matrix in Eq.(\ref{gauges}), we may draw the four diagrams in Fig. 1 which allow for non diagonal entries in the mass matrix for the down
quark sector of the form ($\Delta_D   D_L d_R + h.c.$) and ($\Delta_d d_L D_R + h.c.$) respectively, which in turn produce a radiative mass for the ordinary down quark. Notice that due to the presence of $K^{0\mu}_R$ in the graphs, mass entries of the form $d_Ld_R$ and $D_LD_R$ are not present.
\begin{center}
\begin{picture}(240,250)(0,0)
\Vertex(60,30){1.5} \Vertex(180,30){1.5}
\ArrowLine(0,30)(60,30)
\ArrowLine(60,30)(120,30)
\ArrowLine(120,30)(180,30)
\ArrowLine(180,30)(240,30)
\PhotonArc(120,30)(60,0,180){4}{8.5}
\Text(120,94)[]{\Large$\times$}
\Text(120,32)[]{\large$\otimes$}
\Text(55,85)[]{\small$K^0_{\mu R}$}
\Text(190,85)[]{\small$Z_\mu^0 (Z_\mu^{'0})$}
\Text(120,98)[b]{\small$\epsilon (\epsilon^{'})$}
\Text(0,20)[lt]{\small$d_L$}
\Text(90,15)[]{\small$D_L$}
\Text(120,20)[]{\small$h_DV$}
\Text(150,15)[]{\small$D_R$}
\Text(240,20)[rt]{\small$D_R$}
\Text(60,20)[]{\small$\alpha^\mu$}
\Text(180,20)[]{\small$\beta^\mu(\beta^{'\mu})$}
\Text(120,140)[]{\small (a)}
\Vertex(60,170){1.5} \Vertex(180,170){1.5}
\ArrowLine(0,170)(60,170)
\ArrowLine(60,170)(120,170)
\ArrowLine(120,170)(180,170)
\ArrowLine(180,170)(240,170)
\PhotonArc(120,170)(60,0,180){4}{8.5}
\Text(120,235)[]{\Large$\times$}
\Text(120,172)[]{\large$\otimes$}
\Text(55,225)[]{\small$K^0_{\mu R}$}
\Text(190,225)[]{\small$Z_\mu^0 (Z_\mu^{'0})$}
\Text(120,240)[b]{\small$\epsilon (\epsilon^{'})$}
\Text(0,160)[lt]{\small$D_L$}
\Text(90,155)[]{\small$d_L$}
\Text(120,160)[]{\small$h_dv_1$}
\Text(150,155)[]{\small$d_R$}
\Text(240,160)[rt]{\small$d_R$}
\Text(60,160)[]{\small$\alpha^\mu$}
\Text(180,160)[]{\small$\beta^\mu(\beta^{'\mu})$}
\Text(120,0)[]{\small (b)}
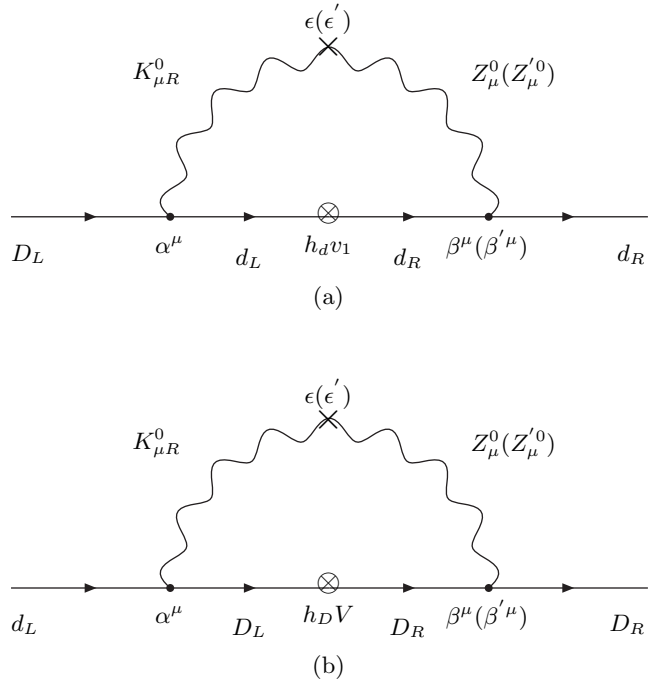
\end{picture} \\ {\small Fig.1. Four one-loop diagrams contributing to the 
radiative generation of the ordinary down quark mass. For the meaning of 
$\alpha$, $\beta$ and $\epsilon$ see the main text.}
\end{center}

The equations in this paper imply for the diagrams in Fig. 1 that: 
$\alpha_\mu = g\gamma_\mu/2$, $\beta_\mu=g\gamma_\mu S_WT_W/3$,
$\beta^\prime_\mu = -g^\prime \gamma_\mu T_W/\sqrt{27}$, 
$\epsilon=-C_Wv_1V$ and $\epsilon^\prime = C_Wv_1V/\sqrt{4C_W^2-1}$.

In a similar way we achieve radiative masses for the charged lepton and
for the Majorana neutrino. The detailed analysis for these leptons will be
presented elsewhere.


\begin{thebibliography}{99}
\bibitem{sm}
For an excellent compendium of the SM see: J.F. Donoghue, E. Golowich, and
B. Holstein, {\it Dynamics of the Standard Model}, (Cambridge University
Press, Cambridge, England, 1992).

\bibitem{pf}
F. Pisano and V. Pleitez, Phys. Rev. D{\bf 46}, 410 (1992);  P.H. Frampton,
Phys. Rev. Lett. {\bf 69}, 2887 (1992); J.C. Montero, F. Pisano and V.
Pleitez, Phys. Rev. D{\bf 47}, 2918 (1993);R. Foot, O.F. Hernandez,
F. Pisano and V. Pleitez, Phys. Rev. D{\bf 47}, 4158 (1993); V. Pleitez
and M.D. Tonasse, Phys. Rev. D{\bf 48}, 2353 (1993); {\it ibid} 5274
(1993); D. Ng, Phys. Rev. D{\bf 49}, 4805 (1994); L. Epele, H. Fanchiotti,
C. Garc\'\i a Canal and D. G\'omez Dumm, Phys. Lett. {\bf 343B}, 291 (1995); M. \"Ozer, Phys. Rev. D{\bf 54}, 4561 (1996).

\bibitem{scalar}
M.D. Tonasse, Phys. Lett. {\bf 381B}, 191 (1996); D. G\'omez Dumm, Int. J. Mod. Phys. A{\bf 11}, 887 (1996); H.N. Long, Mod. Phys. Lett. {\bf A13}, 1865 (1998) [hep-ph/9711204]; N.T. Anh, N.A. Ky, and H.N. Long, Int. J. Mod. Phys. A{\bf 15}, 283 (2000) [hep-ph/9810273]; {\it ibid}, A{\bf 16}, 541 (2001) [hep-ph/0011201].

\bibitem{valle}
M. Singer, J.W.F. Valle and J. Schechter, Phys. Rev. D{\bf 22}, 738 (1980);
R. Foot, H.N. Long and T.A. Tran, Phys. Rev. D{\bf 50}, R34 (1994); H.N.
Long, Phys. Rev. D{\bf 53}, 437 (1996); {\it ibid} D{\bf 54}, 4691 (1996);
V. Pleitez, Phys. Rev. D{\bf 53}, 514 (1996).

\bibitem{ozer}
M. \"Ozer, Phys. Rev. D{\bf 54},1143(1996); T. Kitabayashi, Phys. Rev. D{\bf 64}, 057301 (2001).

\bibitem{spm}
L.A. S\'anchez, W.A. Ponce, and R. Mart\'\i nez, Phys. Rev. D{\bf
64}, 075013 (2001) [hep-ph/0103244].

\bibitem{mps}
R. Mart\'\i nez, W.A. Ponce and L.A. S\'anchez, Phys. Rev. D{\bf 65},
055013 (2002) [hep-ph/0110246].

\bibitem{pfs}
W.A. Ponce, J.B. Fl\'orez, and L.A. S\'anchez, Int. J. Mod. Phys. A{\bf 17}, 643 (2002) [hep-ph/0103100].

\bibitem{pgs}
W.A. Ponce, Y. Giraldo and L.A. S\'anchez, ``{\it Systematic study of 331
models}'', in proceedings of the VIII Mexican Workshop of Particles and
Fields, Zacatecas, Mexico, 2001. Edited by J.L. D\'\i az-Cruz {\it et al.} (AIP Conf. Proceed. Vol. {\bf 623}, N.Y., 2002), pp. 341-346 [hep-ph/0201133].

\bibitem{goldstone}
J. Goldstone, A. Salam and S. Weinberg, Phys. Rev. {\bf 127}, 965 (1962);
S. Bludman and A. Klein, {\it ibid}, {\bf 131}, 2364 (1963); W. Gilbert,
Phys. Rev. Lett. {\bf 12}, 713 (1964).

\bibitem{langa}
This result is taken from the 2001 update of the contribution {\it
Electroweak Model and Constraints on New Physics} by J. Erler and P.
Langacker, in Ref. \cite{pdb} (available at http://pdg.lbl.gov/).

\bibitem{pdb}
D.E. Groom {\it et al.}, Eur. Phys. J. C{\bf 15}, 1 (2000).


\end{thebibliography}
\end{document}